\documentclass[prb,showpacs,twocolumn,byrevtex]{revtex4}
\usepackage{graphicx}
\usepackage{float}
\usepackage{bm}
\begin{document}

\bibliographystyle{apsrev}

\preprint{Draft version, not for distribution}

%
%

\title[fete]{Normal state charge dynamics of
Fe$_{1.06}$Te$_{0.88}$S$_{0.14}$ superconductor probed with infrared
spectroscopy}

%
%

\author{N. Stojilovic$^1$, A. Koncz$^2$, L.W. Kohlman$^2$, Rongwei Hu$^{3, \dag}$,
C. Petrovic$^3$ and S.V. Dordevic$^{2, \ast}$}

\affiliation{$^1$Department of Physics and Astronomy, University of
Wisconsin Oshkosh, Oshkosh, WI 54901, USA}

\affiliation{$^2$Department of Physics, The University of Akron,
Akron, OH 44325, USA}

\affiliation{$^3$Condensed Matter Physics and Materials Science
Department, Brookhaven National Laboratory, Upton, NY 11973, USA}

\date{\today}

%
%
\begin{abstract}
We have used optical spectroscopy to probe the normal state
electrodynamic response of Fe$_{1.06}$Te$_{0.88}$S$_{0.14}$, a member
of the 11 family of iron-based superconductors with T$_c$= 8~K.
Measurements have been conducted over a wide frequency range (50 -
50000~cm$^{-1}$) at selected temperatures between 10 and 300 K. At
low temperatures the material behaves as an "incoherent metal": a
Drude-like peak is absent from the optical conductivity, and all
optical functions reveal that quasiparticles are not well defined
down to the lowest measured temperature. We introduce "generalized
spectral weight" analysis and use it to track temperature induced
redistribution of spectral weight. Our results, combined with
previous reports, indicate that the 11 family of iron-based
superconductors might be different from other families.
\end{abstract}

%
%
%
%
%

\maketitle

\section{Introduction}

Iron-based superconductors are currently at the focus of
condensed-matter research. Discovered less than two years ago
\cite{Kamihara08}, these novel materials have attracted attention not
only because of their high critical temperatures, but also because of
their similarities with cuprates. Their phase diagram resembles that
of the cuprates, most notably, the superconducting state seems to
develop from an unconventional normal state, after magnetic order is
destroyed by doping. They are also layered materials, consisting of
FeAs, FeTe, FeS or FeSe planes, separated by spacer layers. However,
there are also some important differences. Most notably, the parent
compounds of cuprates are antiferromagnetic insulators, whereas the
parent compounds of iron-based superconductors are antiferromagnetic
spin density wave (SDW) metals \cite{norman08}.

Optical spectroscopy is a powerful probe of electrodynamic response
of high-T$_c$ superconductors \cite{Basov05,Dordevic06,wang09}.
Optical constants provide insight into low-energy excitations and
charge dynamics, critical for understanding physics of strongly
correlated systems. The information obtained from optical constants
can be used to test existing theories and/or stimulate development of
new theoretical models. Optical spectroscopy is also a crucial
experimental method for electronic band structure determination.

Several families of iron-based superconductors have been discovered
and they are conveniently refereed to as the "11", "1111" or "122"
families \cite{wang09}. The 11 family is peculiar because the spacer
layers are absent, and it is believed that this family will allow the
intrinsic properties of iron-containing  planes to be isolated.
However, most of optical studies so far have focused on the 122
family \cite{yang08,hu08,Pfuner09,akrap09,kim09,heumen09,gorshunov09,wu09}
(and to a lesser extent 1111 family \cite{qazilbash08,chen091111})
for which large single crystals can be grown. In this work, we have
investigated the electrodynamic response of a member of the 11 (FeTe)
family. To the best of our knowledge this is the first optical study
of a superconducting member of this family. The only previous IR
study on 11 family was on a non-superconducting Fe$_{1.05}$Te
(Ref.~\onlinecite{Chen09}).

Structural analysis has shown that the exact chemical composition of
the analyzed sample is Fe$_{1.06}$Te$_{0.88}$S$_{0.14}$
(Ref.~\onlinecite{petrovic09}). Note that in addition to being doped
with S, this sample also has 6~$\%$ of excess iron, which might play
important role in charge dynamics \cite{Chen09,zhang09}.
Magnetization measurements have revealed that the studied system
undergoes structural and magnetic transitions, with transition
temperature around 23 K (Ref.~\onlinecite{petrovic09}). Transport
measurements on the other hand do not display any signatures of these
transitions, and DC resistivity monotonically increases as
temperature decreases down to 8 K, when the system undergoes
superconducting transition and the resistivity abruptly drops to
zero.

Infrared (IR) reflectance measurements were performed at The
University of Akron on a Bruker IFS 66v/s, whereas UV-visible
experiments were conducted using Varian/Cary 300. An overfilling
technique was used to obtain the absolute values of reflectance from
the sample with surface area of approximately 1~mm $\times$ 1~mm
(Ref.~\onlinecite{Homes93}). Electrodynamic response was probed in
the frequency range 50 - 50 000~cm$^{-1}$ (6 meV - 6.2 eV) and as a
function of temperature in the range 10~K - 300~K, all in the normal
state. The optical constants were extracted from reflectance data
using Kramers-Kronig (KK) analysis. In addition, we have performed
magneto-optical measurements at 4.2 K (in the superconducting state)
with magnetic fields up to 18 Tesla.

\section{Experimental results}

\begin{figure}[t]
\vspace*{-1.0cm}%
\centerline{\includegraphics[width=9.5cm]{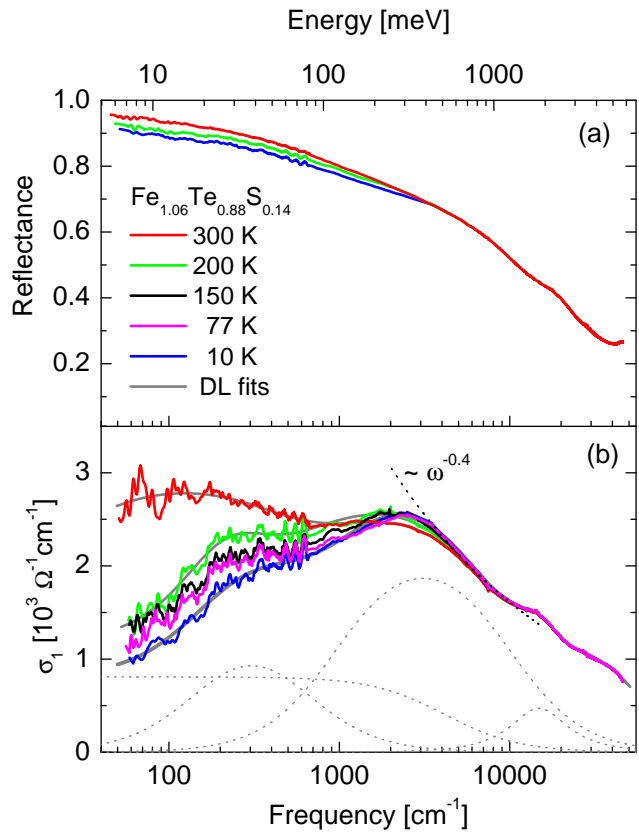}}%
\vspace*{-1.0cm}%
\caption{(Color online). (a) Raw reflectance data of iron-based
superconductor Fe$_{1.06}$Te$_{0.88}$S$_{0.14}$ at several different
temperatures ranging from 10 K to room temperature. The temperature
dependence is restricted to the region below about
4000~cm$^{-1}$. (b) The optical conductivity
$\sigma_1(\omega)$ extracted directly from reflectance using KK
analysis. The total DL fit, as well as the individual components
of the fit, are shown with gray lines. The dashed black line is the fit
to the power-law behavior at higher frequencies.}
\vspace*{-0.5cm}%
\label{fig:ref}
\end{figure}

Figure \ref{fig:ref} displays raw reflectance data R($\omega$) and
the real part of optical conductivity $\sigma_1(\omega)$. The
absolute value of reflectance gradually decreases with frequency from
about 0.9 at 50~cm$^{-1}$ to about 0.25 at 50000~cm$^{-1}$, which can
be interpreted as metallic behavior. However the temperature
dependence is opposite from expected \cite{dressel-book}, as the
reflectivity decreases with decreasing temperature. This anomalous
behavior is even more obvious in $\sigma_1(\omega)$: zero-energy
(Drude-like) peak is absent from the data (except maybe at 300 K) and
the conductivity decreases monotonically with decreasing temperature
at the lowest measured frequencies. This result indicates that at low
temperatures quasiparticles are not well defined, which will become
even more obvious from the extended Drude analysis below.
Fe$_{1.06}$Te$_{0.88}$S$_{0.14}$ therefore can be considered to be an
"incoherent metal" \cite{craco09}.

Although the overall behavior of reflectance is metallic, the plasma
edge cannot be clearly resolved in the spectra, as reflectance
gradually decreases with frequency. At higher frequencies the
response is dominated by interband transitions, but they also are not
easily discerned in the spectra. A shoulder in optical conductivity
around 14000~cm$^{-1}$ (1.7 eV) may originate from transitions
involving iron {\it 3d} states, similar to what was predicted by
Haule et al. in a theoretical study of another iron-based
superconductor \cite{Haule08}. We also do not observe any phonon
peaks in the far-IR, similar to Fe$_{1.05}$Te
(Ref.~\onlinecite{Chen09}). Structural and magnetic transitions at
23~K do not seem to have any significant effect on optical spectra.

Theoretical calculations of FeSe, based on LDA+DMFT, predicted
"incoherent metal" with a pseudogap at low frequencies
\cite{craco09}, in accord with our results. These calculations also
predict a smooth crossover of optical conductivity to a power law
behavior $\simeq \omega ^{- \eta}$ at higher frequencies
\cite{craco09}. Dashed black line in Fig.~\ref{fig:ref}(b) represents
the best fit to the optical conductivity and the obtained power law
$\eta \simeq 0.4$.


To gain further insight into the electronic properties of
Fe$_{1.06}$Te$_{0.88}$S$_{0.14}$ we fit the data using a standard
Drude-Lorentz (DL) model \cite{dressel-book,Basov05,Dordevic06}. The
minimal model to achieve a good fit consisted of a Drude and three
Lorentzian modes, centered at around 300~cm$^{-1}$, 3000~cm$^{-1}$
and 14800~cm$^{-1}$. The total fits at all temperatures, as well as
the three individual contributions at 10~K, are shown in
Fig.~\ref{fig:ref}(b) with gray lines. The lowest lying oscillator
displays most prominent temperature dependence. Its energy and
intensity grow significantly as temperature decreases. The mid-IR
peak at 3000~cm$^{-1}$ (372~meV) might be a generic feature of
iron-based superconductors; similar peaks have been observed in other
families \cite{wang09}. The oscillator at 14800~cm$^{-1}$ simulates
the effect of interband transitions, presumably involving iron {\it
3d} states, as discussed above.

\begin{figure}[t]
\vspace*{-1.0cm}%
\centerline{\includegraphics[width=9.5cm]{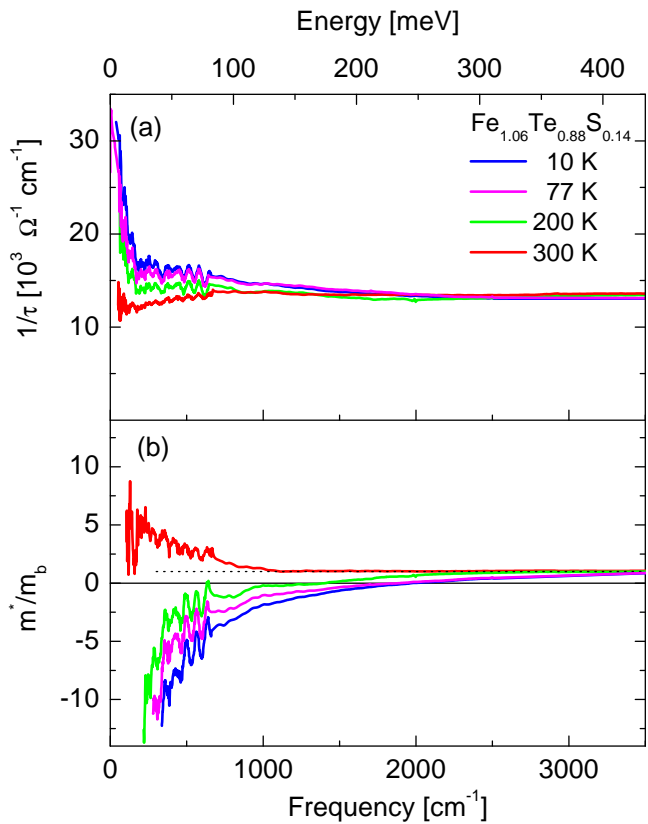}}%
\vspace*{-1.0cm}%
\caption{(Color online). Optical scattering rate
$1/\tau(\omega)$ and effective mass $m^*(\omega)/m_b$
obtained from Eqs.~\ref{eq:tau} and \ref{eq:mass}. Both optical
functions reveal the failure of the quasipaticle concept at
low temperatures.}%
\vspace*{-0.5cm}%
\label{fig:tau}
\end{figure}

In the one-component approach one assumes that only a single type of
carriers are present in the system, but their scattering rate
acquires frequency dependence \cite{Basov05,Dordevic06,dressel-book}.
Within the so-called "extended" Drude model one calculates the
optical scattering rate $1/\tau(\omega)$ and effective mass
$m^*(\omega)/m_b$ from the complex optical conductivity
$\sigma(\omega)$ as:

\begin{equation}
\frac{1}{\tau(\omega)}=\frac{\omega_{p}^{2}}{4 \pi} \Re \Big[
\frac{1} {\sigma(\omega)} \Big]
\label{eq:tau}%
\end{equation}

\begin{equation}
\frac{m^{*}(\omega)}{m_b} = \frac{\omega_{p}^{2}}{4 \pi} \Im
\Big[ \frac{1} {\sigma(\omega)} \Big] \frac{1}{\omega}
\label{eq:mass}%
\end{equation}
where the plasma frequency $\omega_p^2=4 \pi e^2 n/m_b$ ($n$ is the
carrier density and $m_b$ their band mass) can be estimated from the
integration of $\sigma_{1}(\omega)$ up to the frequency of the onset
of interband absorption. However, as pointed out above, the plasma
edge is not very prominent in the spectra of
Fe$_{1.06}$Te$_{0.88}$S$_{0.14}$, rendering the value of plasma
frequency ambiguous. Instead, we fit the value of $\omega_p$ in
Eq.~\ref{eq:mass} so that the effective mass at frequencies around
3500~cm$^{-1}$ is equal to unity (dashed line in
Fig.~\ref{fig:tau}(b)). The best fit is achieved for $\omega_p$=
26,000~cm$^{-1}$ (3.2 eV) and the results of the analysis are shown
in Fig.~\ref{fig:tau}. The features of an "incoherent metal" are now
obvious. The scattering rate is relatively flat and featureless at
room temperature, but at low temperatures it develops a peak below
180~cm$^{-1}$. The effective mass, on the other hand, becomes
negative. These feature of the spectra are indications that the
quasiparticle concept, on which the extended Drude model is based, is
not applicable at low temperatures.


\section{Generalized spectral weight analysis}

\begin{figure}[t]
\vspace*{2.0cm}%
\centerline{\includegraphics[width=9.5cm]{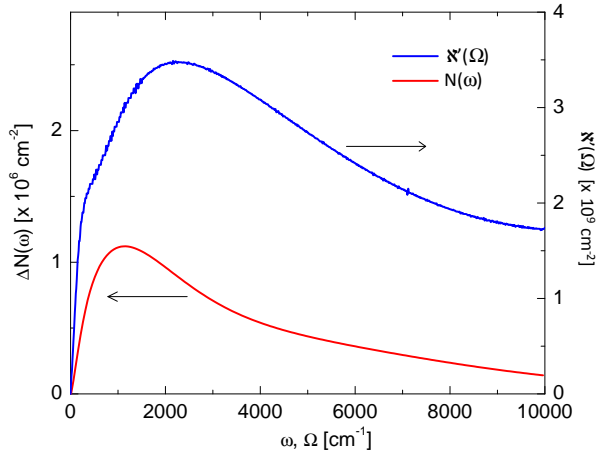}}%
\vspace*{-2.5cm}%
\caption{(Color online). Spectral weight (Eq.~\ref{eq:delta-n})
and generalized spectral weight (Eq.~\ref{eq:aleph}) analysis for
Fe$_{1.06}$Te$_{0.88}$S$_{0.14}$ at T$_1$ = 10 K and T$_2$ = 77 K.}%
\vspace*{-0.5cm}%
\label{fig:gsw}
\end{figure}

Model-independent sum rules are important tools in condensed-matter
physics \cite{mahan-book}. The so-called effective spectral weight
function N($\omega$) is frequently used for the analysis of IR
spectra. It is defined as \cite{dressel-book,Basov05}:

\begin{equation}
N(\omega)=\int_0^{\omega}\sigma_1(x)dx,
\label{eq:n}%
\end{equation}
and for $\omega \rightarrow \infty$ it becomes the global oscillator
strength sum rule:

\begin{equation}
N(\infty)=\int_0^{\infty}\sigma_1(x)dx=\frac{\pi n e^2}{2 m_e},
\label{eq:sumrule}%
\end{equation}
which is a statement on the conservation of electric charge
\cite{mahan-book}. Eq.~\ref{eq:n} is often used to quantify spectral
weight redistribution between temperatures T$_1$ and T$_2$ in the
form:

\begin{equation}
\Delta N(\omega)=N_{T_1}(\omega) - N_{T_2}(\omega).
\label{eq:delta-n}%
\end{equation}
Fig.~\ref{fig:gsw} shows the results of this analysis applied to
Fe$_{1.06}$Te$_{0.88}$S$_{0.14}$ at T$_1$= 77~K and T$_2$= 10~K.
$\Delta N(\omega)$ has a characteristic shape, which indicates that
the spectral weight is removed from the low frequency region, below
1000~cm$^{-1}$ and is transferred to higher frequencies, in the
region around 2000-4000~cm$^{-1}$. Within the error bars, the
spectral weigh is recovered by the mid-IR region. However, $\Delta
N(\omega)$ spectrum cannot reveal the energy scale at which the
transfer of spectral weight occurs.

In order to address this question we introduce "generalized spectral
weight" function $\aleph(\Omega)$:

\begin{equation}
\aleph(\Omega)=\int_0^{\infty}\sigma_1^{T_1}(x+\Omega)\sigma_1^{T_2}(x)dx
\label{eq:aleph}%
\end{equation}
The idea comes from the correlation functions frequently used in
signal processing \cite{proakis-book}. More recently
(auto)correlation function was used for the analysis of ARPES
spectra, where it is directly related to the quasiparticle density of
states \cite{chatterjee06}. Note that $\aleph(\Omega)$ is a function
of the frequency shift $\Omega$, not the upper integration limit
$\omega$. Function $\aleph(\Omega)$ is expected to display
characteristic features at the values of energy shifts $\hbar \Omega$
that connect regions between which a large amount of spectral weight
is transferred. In practical applications, function $\aleph(\Omega)$
is usually dominated by the spectral weight which does not
participate in redistribution, and these characteristic features
might not be obvious. In those cases the first derivative of
$\aleph(\Omega)$ is useful. Examples of these calculations will be
presented in a separate publication \cite{dordevic10}.

In Fig.~\ref{fig:gsw} we show the results of generalized spectral
weight analysis (Eq.~\ref{eq:aleph}) applied to
Fe$_{1.06}$Te$_{0.88}$S$_{0.14}$ at T$_1$ = 77~K and T$_2$ = 10~K.
The $\aleph(\Omega)$ is dominated by the spectral weight that does
not participate in transfer, so instead we display the first
derivative $\aleph'(\Omega)$. The $\aleph'(\Omega)$ spectrum reveals
a broad peak centered around 2250~cm$^{-1}$ (280~meV). We take
this as the characteristic energy scale over which the majority of
spectral weight is transferred between 10 and 77~K.


\section{Discussion}

In a density functional study of FeS, FeSe and FeTe it was reported
that the Fermi surface and electronic structure of these compounds is
similar to those of iron pnictides (1111 and 122 families)
\cite{Subedi08}. Therefore, the 11 family was suppose to be a model
system in which to study the intrinsic properties of iron-containing
planes. However, our results, combined with previous IR studies,
indicate that there might be some important differences between the
electronic structure of 11 and the other families.

The reflectance of Fe$_{1.06}$Te$_{0.88}$S$_{0.14}$ is similar to the
reflectance of Fe$_{1.05}$Te (Ref.~\onlinecite{Chen09}). In the
far-IR region the reflectance of Fe$_{1.05}$Te decreases with
decreasing temperature, resulting in a reduction of conductivity in
far-IR region, similar to what we observe in
Fe$_{1.06}$Te$_{0.88}$S$_{0.14}$. Chen et al. speculate that this
incoherent transport in Fe$_{1.05}$Te is caused by strong scattering
from excess iron \cite{Chen09}. However, there are also some
important differences between Fe$_{1.06}$Te$_{0.88}$S$_{0.14}$ and
Fe$_{1.05}$Te. We observe no rapid increase in conductivity at low
frequencies for 10 K measurements. On the other hand in Fe$_{1.05}$Te
a narrow Drude-like peak develops in the optical conductivity at low
temperatures. This coherent behavior appears below structural and
magnetic phase transition at 65~K, which implies that it is related
to SDW order. In Fe$_{1.05}$Te this phase transition has stronger
influence on charge transport than in
Fe$_{1.06}$Te$_{0.88}$S$_{0.14}$: the DC resistivity changes
character from insulating to metallic below the transition.

Incoherent charge transport in the 11 family should be contrasted
with a coherent response which has been observed in 1111
\cite{qazilbash08,chen091111} and 122 families
\cite{yang08,hu08,Pfuner09,akrap09,kim09,heumen09,gorshunov09}.
Infrared spectra of both undoped (parent) and doped phases of these
families display well defined Drude-like modes. Similar to
Fe$_{1.05}$Te SDW transition has a dramatic effects on their optical
properties. The response of parent compounds BaFe$_2$As$_2$ and
SrFe$_2$As$_2$ becomes even more coherent below the SDW transitions
\cite{hu08} as the width of Drude mode is reduced by an order of
magnitude. The infrared spectra of these parent compounds are
dominated by the mid-IR peak, which may have the same origin as the
peak we observe in Fe$_{1.06}$Te$_{0.88}$S$_{0.14}$ around
3000~cm$^{-1}$ (372~meV).

The absence of SDW gap from IR spectra of both
Fe$_{1.06}$Te$_{0.88}$S$_{0.14}$ and Fe$_{1.05}$Te
(Ref.~\onlinecite{Chen09}) is also interesting. Recent ARPES study of
a parent compound Fe$_{1+x}$Te (Ref.~\onlinecite{Xia09}) has also
revealed that the SDW gap is absent. Optical spectra reveal that
spectral weight is shifting with temperature, and the generalized
spectral weight analysis we introduced indicates that a typical
energy scale for the shift is about 280~meV. The spectral weight of
Fe$_{1.06}$Te$_{0.88}$S$_{0.14}$ is removed from the low-energy
region, which can be interpreted as a pseudogap feature, however we
point out that this behavior starts already at room temperature
(Fig.~\ref{fig:ref}) and therefore is unlikely to be related to SDW
transition. All this indicates that the electronic structure of the 11
family might be different from the 1111 and 122 families, for which clear
signatures of SDW gaps have been observed \cite{wang09}. IR studies
on these two families have found a gap (or even several gaps) in the
excitation spectra. On the other hand both the parent compound
\cite{Chen09} and a doped sample studied in this work did not reveal
the presence of a gap in their excitation spectra.

\begin{figure}[t]
\vspace*{2.0cm}%
\centerline{\includegraphics[width=9.5cm]{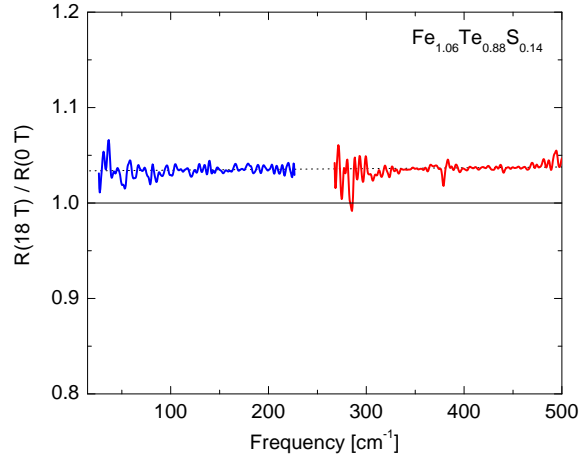}}%
\vspace*{-2.5cm}%
\caption{(Color online). Magneto-optical ratio R(18 T)/ R(0 T)
obtained at 4.2~K. We found no deviation from straight line, which
indicates that magnetic field does not affect the optical properties.
The measurement system causes vanishingly small signal intensity in
the region around 250~cm$^{-1}$, and this part of the spectrum is not
shown \cite{dordevic09}.}%
\vspace*{-0.5cm}%
\label{fig:magnet}
\end{figure}

Finally we address what happens below 8~K, when the system becomes
superconducting. We have performed magneto-optical studies in 18
Tesla superconducting magnet at the National High Magnetic Field Lab.
Fig.~\ref{fig:magnet} displays the results of these measurements. The
magneto-reflection ratio R(18 T)/R(0 T) is shown as a function of
frequency. Apart from the vertical offset \cite{dordevic09}, within
the noise level the ratio is a straight line, which indicates the
absence of field-induced effects in Fe$_{1.06}$Te$_{0.88}$S$_{0.14}$.
This is in contrast with BaFe$_{2-x}$Co$_{x}$As$_2$ with T$_c$=~22~K
(optimally doped member of the 122 family), where clear field-induced
changes have been observed caused by the suppression of the
superconducting gap \cite{schafgans09}. We speculate that
field-induced changes in Fe$_{1.06}$Te$_{0.88}$S$_{0.14}$ are not
observed because either they are below the detection limit of our
experiment, or because the superconductiong gap is outside of our
frequency window.


\section{Summary}

In summary, we have presented the results of infrared and optical
spectroscopy studies of novel iron-based superconductor
Fe$_{1.06}$Te$_{0.88}$S$_{0.14}$. The results indicate incoherent
normal state charge transport and absence of well defined
quasiparticles at all temperatures down to T$_c$. We have introduced
"generalized spectral weight analysis" and used it to track
redistribution with temperature. The analysis reveals that the
characteristic energy scale for the spectral weight shifts is
approximately 280~meV. Our results, combined with previous reports,
indicate that there are important differences between 11 and other
families of iron-based superconductors.


We thank D.N. Basov and C.C. Homes for critical reading of the
manuscript. Special thanks to R.D. Ramsier for the use of his
equipment. Research at Brookhaven National Laboratory (Rongwei Hu and
C. P.) is supported by the U.S. Department of Energy, Office of Basic
Energy Sciences, Division of Materials Sciences and Engineering under
Award $\#$ (DE-Ac02-98CH10886).

%
%

\begin{references}

\bibitem{Kamihara08} Y. Kamihara, T. Watanabe, M. Hirano and H. Hosono,
    Journal of the American Chemical Society {\bf 130}, 3296 (2008).

\bibitem{norman08} M.R. Norman, Physics {\bf 1}, 21 (2008).

\bibitem{Basov05} D.N. Basov and T. Timusk, Reviews of Modern Physics {\bf 77},
    721 (2005).

\bibitem{Dordevic06} S.V. Dordevic and D.N. Basov, Annalen der Physik {\bf 15},
    545 (2006).

\bibitem{wang09} W.Z. Hu, Q.M. Zhang and N.L. Wang, Physica C {\bf
    469}, 545 (2009) 

\bibitem{yang08} J.Yang, D. Huvonen, U. Nagel, T. Room, N. Ni, P.C.
    Canfield, S.L. Budko, J.P. Carbotte and T. Timusk, Physical Review
    Letters {\bf 102}, 187003 (2009).

\bibitem{hu08} W.Z. Hu, J. Dong, G. Li, Z. Li, P. Zheng, G.F. Chen,
    J.L. Luo and N.L. Wang, Physical Review Letters {\bf 101}, 257005 (2008). 

\bibitem{Pfuner09} F. Pfuner, J.G. Analytis, J.-H. Chu, I.R. Fisher
    and L. Degiorgi, The European Physical Journal B {\bf 67} 513
    (2009). 

\bibitem{akrap09} A. Akrap, J.J. Tu, L.J. Li, G.H. Cao, Z.A. Xu and C.C.
    Homes, Physical Review B {\bf 80}, 180502(R) (2009)   

\bibitem{kim09} K.W. Kim, M. Rossle, A. Dubroka, V.K. Malik, T. Wolf and
C. Bernhard, arXiv:0912:0140 (2009). 

\bibitem{heumen09} E. van Heumen, Y. Huang, S. de Jong, A.B.
    Kuzmenko, M.S. Golden and D. van der Marel arXiv:0912.0636v1 (2009). 

\bibitem{gorshunov09} B. Gorshunov, D. Wu, A. A. Voronkov, P. Kallina,
    K. Iida, S. Haindl, F. Kurth, L. Schultz, B. Holzapfel and M. Dressel,
    arXiv:0912.1256v1 (2009). 

\bibitem{wu09} D. Wu, N. Barisic, P. Kallina, A. Faridian, B. Gorshunov,
N. Drichko, L. J. Li, X. Lin, G. H. Cao, Z. A. Xu N. L. Wang and M. Dressel,
arXiv:0912.3334v1. 

\bibitem{qazilbash08} M.M. Qazilbash, J.J. Hamlin, R.E. Baumbach,
    M.B. Maple and D.N. Basov, arXiv:0808:3748 (2009). 

\bibitem{chen091111} Z.G. Chen, R.H. Yuan, T. Dong and N.L. Wang,
    arXiv:0910:1318v1 (2009). 

\bibitem{Chen09} G.F. Chen, Z.G. Chen, J. Dong, W.Z. Hu, G. Li, X.D.
Zhang, P. Zheng, J.L. Luo and N.L. Wang, Physical Review B.
{\bf 79}, 140509(R) (2009). 

\bibitem{petrovic09} Rongwei Hu, Emil S. Bozin, J.B. Warren and C. Petrovic, 	
Physical Review B {\bf 80}, 214514 (2009).

\bibitem{zhang09} Lijun Zhang, D. J. Singh, and M. H. Du,
Physical Review B {\bf 79}, 012506 (2009). 

\bibitem{Homes93} C.C. Homes, M.A. Reedyk, D.A. Crandles and T. Timusk,
    Applied Optics {\bf 32}, 2976 (1993).

\bibitem{dressel-book} M. Dressel and G. Gruner, {\it Electrodynamics
    of Solids}, Cambridge University Press, Cambridge (2001).

\bibitem{craco09} L. Craco, M.S. Laad and S. Leoni, arXiv:0910.3828v1 (2009).

\bibitem{Haule08} K. Haule, J.H. Shim and G. Kotliar, Physical Review Letters
    {\bf 100}, 226402 (2008).

\bibitem{mahan-book} G.D. Mahan {\it Many-particle physics}, Plenum (1993).

\bibitem{proakis-book} J.G. Proakis, {\it Digital Signal Processing},
    4$^{th}$ edition, New Delhi, Prentice-Hall of India (2007).

\bibitem{chatterjee06} U. Chatterjee, M. Shi, A. Kaminski, A.
    Kanigel, H. M. Fretwell, K. Terashima, T. Takahashi, S.
    Rosenkranz, Z. Z. Li, H. Raffy, A. Santander-Syro, K. Kadowaki,
    M. R. Norman, M. Randeria and J. C. Campuzano, Physical Review Letters
    {\bf 96}, 107006 (2006).

\bibitem{dordevic10} S.V. Dordevic et al. unpublished.

\bibitem{Subedi08} A. Subedi, L. Zhang, D.J. Singh and M.H. Du,
    Physical Review B {\bf 78}, 134514 (2008).

\bibitem{Xia09} Y. Xia, D. Qian, L. Wray, D. Hsieh, G.F. Chen, J.L.
    Luo, N.L. Wang and M.Z. Hasan, Physical Review Letters {\bf 103},
    037002 (2009). 

\bibitem{dordevic09} S.V. Dordevic, L.W. Kohlman, L. C. Tung, Y.-J.
    Wang, A. Gozar, G. Logvenov and I. Bozovic, Physical Review B
    {\bf 79}, 134503 (2009).

\bibitem{schafgans09} A.A. Schafgans, D.N. Basov, et. al. unpublished.


\end{references}

$^{\dag}$Present address: Ames Laboratory US DOE and Department of
Physics and Astronomy, Iowa State University, Ames, IA 50011

$^{\ast}$e-mail address: {\it dsasa@uakron.edu}

\end{document}